# Optimization and Simulation of Startup Control for Space Nuclear Power Systems with Closed Brayton Cycle based on NuHeXSys

Chengyuan Li[a,b], Leran Guo[a], Shanfang Huang[a,*], Jian Deng[b]

[a] Department of Engineering Physics, Tsinghua University, Beijing, 100084, China

[b] National Key Laboratory of Nuclear Reactor Technology, Nuclear Power Institute of China, Chengdu, 610213, China

\* Corresponding author: sfhuang@mail.tsinghua.edu.cn

**Highlights**

1. Developed NuHeXSys analysis code to simulate transients of CBC.
2. NSGA-II optimized startup control, reducing time by 1260 seconds.
3. Reduced external energy demand by 17% during startup.

**Abstract**

This paper presents the development and optimization of a Space Nuclear Power System (SNPS) utilizing a helium-xenon gas-cooled Closed Brayton Cycle (CBC). A comprehensive dynamic system analysis code NuHeXSys (Nuclear Helium-Xenon Brayton Cycle Power System) was created, integrating non-ideal gas properties, a multi-channel thermal-hydraulic reactor core, and detailed turbo-machinery components. The innovation lies in applying an evolutionary algorithm (NSGA-II) to optimize the startup control sequence, significantly reducing startup time and energy consumption. Model verification shows parameter deviations within 10%, confirming its accuracy. The optimized control strategy reduced startup time by 1260 seconds and lowered external energy demand by 17%, demonstrating improved efficiency and operational stability for deep space missions. This work provides a foundation for future advancements in optimizing space nuclear power systems.

# 1 Introduction

Deep space exploration missions are becoming increasingly critical in addressing Earth's looming issues such as resource depletion and overpopulation. Missions extending beyond 10 years and requiring power levels exceeding 100 kWe, such as the exploration of Jupiter's icy moons, necessitate advanced power systems. Among these, a nuclear power system with a Closed Brayton Cycle (CBC) and a direct-cycle helium-xenon gas-cooled reactor stands out as an optimal solution[1].

This Space Nuclear Power System (SNPS) benefits from the direct coupling of the power system and reactor core, where He-Xe gas circulates through the reactor to drive the turbine. The integrated Brayton cycle, using He-Xe gas, delivers higher power levels and superior power-to-mass ratios compared to other space reactors. This design also circumvents the thermal transfer limits of heat pipes and minimizes system weight by eliminating intermediate heat extraction devices[2]. Additionally, the He-Xe gas mixture enhances heat removal efficiency and reduces the compressor mass by decreasing compression work and the number of compressor rotor blade stages[3].

The startup phase is crucial for the operational success of the SNPS as power consumption before startup completion is supported by limited-capacity storage batteries. If startup fails, it cannot be repeated, making it imperative to optimize startup control to reduce startup time. Therefore, predicting dynamic behavior and optimizing the control processes of these systems is vital.

A few control methods have been explored in existing research. Skorlygin[4] introduced a startup optimization method that measures temperature, calculates the critical position of the control rod, and adjusts it using bidirectional movement to achieve the desired startup time. Prikot et al.[5] enhanced the control algorithm of the TOPAZ-II space nuclear power system, achieving precise power control and protecting the thermionic emitter from overheating during startup, thus improving startup performance and stability. Zeng et al.[6] used a nonlinear model and fuzzy PID controller to optimize reactivity introduction, power steps, and load-following control of TOPAZ-II. Ma et al.[7] analyzed the startup control of a megawatt heat pipe reactor coupled to a close Brayton cycle. However, the strategies are mainly dependent on expert knowledge. Further information on modeling and control issues in nuclear power systems is available in reference[8].

Heuristic optimization algorithms offer an effective approach for optimizing system control, particularly in complex search spaces where conventional methods are inadequate. For single-objective optimization, algorithms such as Differential Evolution[9], CMA-ES[10], and ISRES[11] can be considered. For optimizing 2 to 3 objectives simultaneously, NSGA-II[12], R-NSGA-II[13], and MOEA/D[14] are suitable choices. For scenarios involving more than three objectives, NSGA-III[15], U-NSGA-III[16], and R-NSGA-III[17] can be utilized.

Optimizing the control needs a system analysis program, but such tools integrating the core and power system is limited. McCann et al.[18] used RELAP-3D to simulate a CBC coupled nuclear reactor, but the transient behavior was not verified. EI-Genk et al.[19] demonstrated the transient process of the S4 reactor coupled with the CBC, but the core modeling lacked detail and did not account for radial thermal-hydraulic variations. Wang et al.[20] utilized a radial multi-channel reactor model in their system analysis but did not address assumptions about gas properties. Geng et al.[21] constructed an analysis tool, but the core model was considered lumped. Ma et al.[22] analyzed the startup performance of the SNPS with multiple Brayton loops, comparing different startup schemes and the impact of space environment temperature, yet their core description was simplified, and neutron dynamics parameters were unspecified. These studies lack of a detailed and integrated system analysis program.

This study develops a system analysis code NuHeXSys (Nuclear Helium-Xenon Brayton Cycle Power System) for modeling the SNPS with the CBC system, which combines the non-ideal He-Xe properties, the multi-channel reactor model, the turbomachinery, the recuperator and the cooler. Based on this modeling, an evolution-based control optimization framework is employed to optimize the control sequence, reducing startup time significantly.

The rest of this paper is structured as follows: Chapter 2 details the modeling of the reactor system and the optimization framework of control optimization. Chapter 3 verifies the model with components, system steady-state and startup transient. Chapter 4 shows results of startup control optimization. Chapter 5 concludes the study and discusses future implications.

## 2 Methodology

This study will first develop a system analysis code NuHeXSys (Nuclear

Helium-Xenon Brayton Cycle Power System) for modeling the closed Brayton cycle reactor system. The non-ideal gas properties of helium-xenon mixtures will be modeled, along with a multi-channel reactor thermal-hydraulic model. Turbomachinery, regenerator, and cooler models within a closed Brayton cycle will also be developed. The Monte Carlo program RMC will be utilized to obtain temperature reactivity coefficients and neutron kinetic parameters. Furthermore, an evolution-based control optimization framework will be employed to optimize the startup control sequence, aiming to reduce startup time. The workflow of this paper is shown in Figure 1.

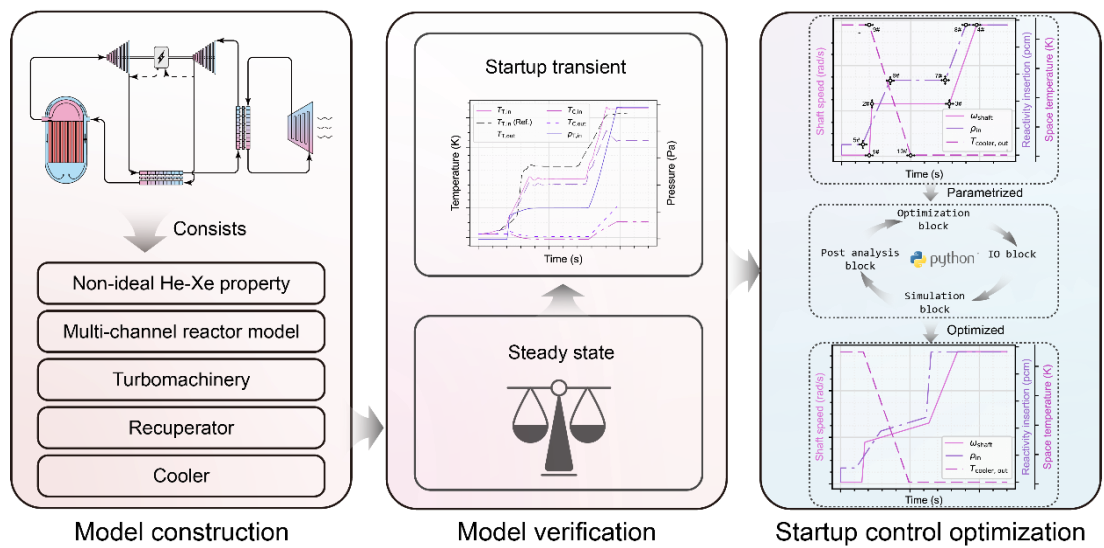

Figure 1 Overview of the workflow

## 2.1 Modeling of the CBC nuclear reactor system

The helium-xenon cooled closed Brayton cycle nuclear reactor system efficiently converts energy through thermodynamic cycles. The helium-xenon gas mixture is compressed, raising its temperature and pressure. Preheated in the regenerator, the gas enters the reactor core, reaching high temperatures. It then expands in the turbine, driving the generator to produce electricity. The expanded gas releases heat in the regenerator and is further cooled in the cooler before returning to the compressor, completing the cycle. Such process is depicted in Figure 2. The block definition diagram reflecting the entities and relationships of the system is shown in Figure 3.

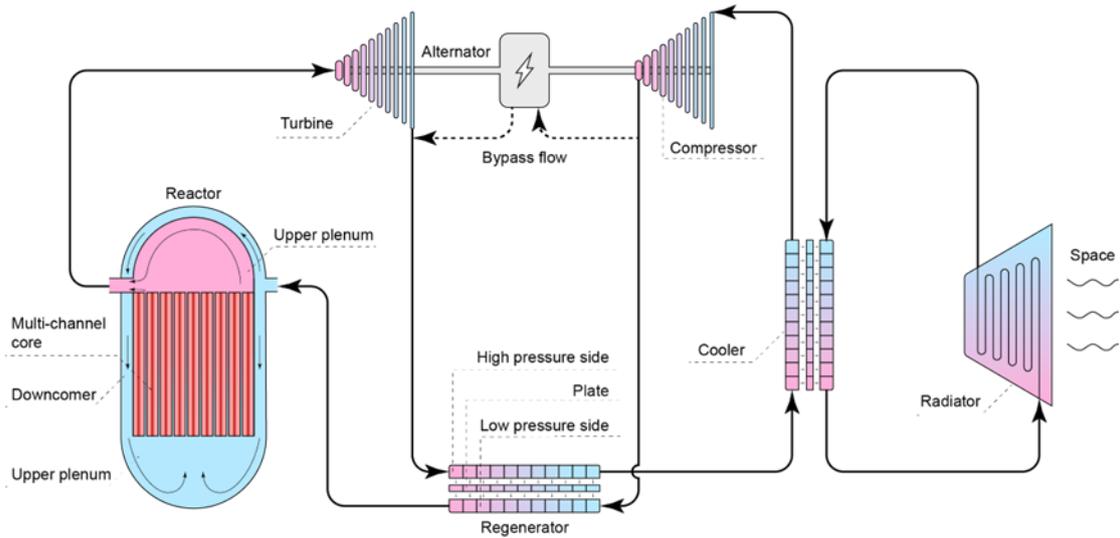

Figure 2 Closed Brayton Cycle reactor system schematic diagram

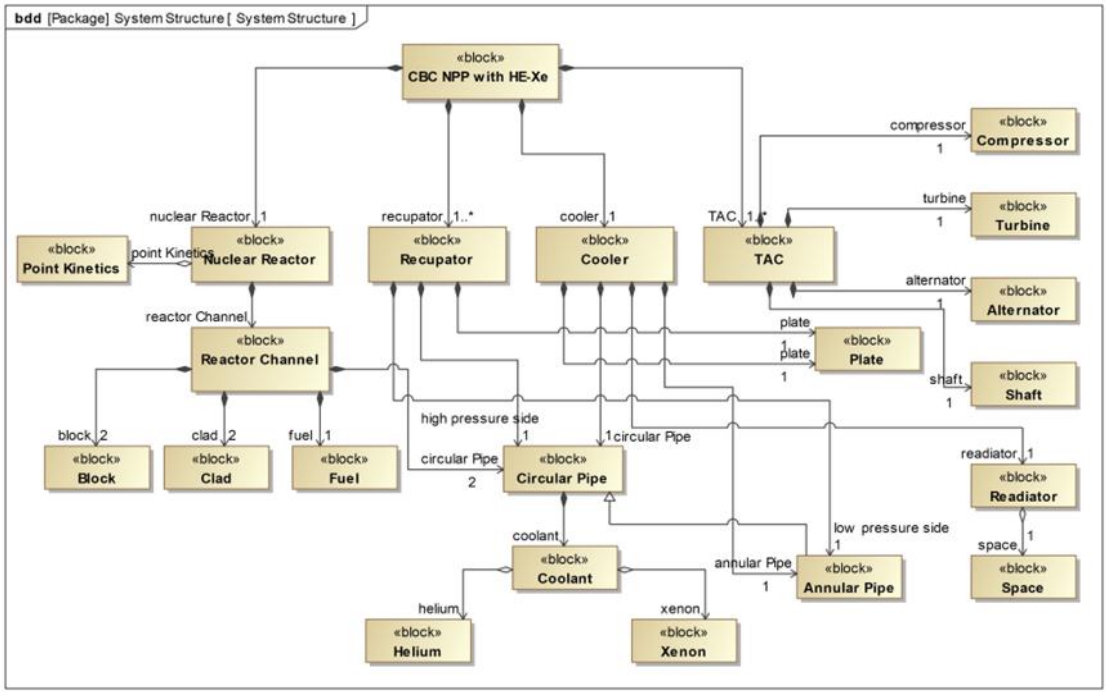

Figure 3 Block definition diagram of closed Brayton Cycle reactor system

### 2.1.1 Non-ideal properties of Helium-Xenon mixture

The ideal gas model assumes no inter-molecular forces and negligible molecular volume. However, in a helium-xenon mixture, xenon's larger molecular volume and significant inter-molecular forces invalidate these assumptions. Non-ideal effects in such mixtures are significantly influenced by xenon proportion, temperature, and pressure, especially at low temperatures, high pressures, and high xenon concentrations. Therefore, it is crucial to consider non-ideal gas effects and select an appropriate equation of state for accurate analysis.

The density calculation utilizes the Virial Equation combined with semi-empirical relationships[23]. This approach has the advantage of accounting for temperature and pressure effects, making it suitable for a wide range of temperature and pressure conditions. However, its accuracy decreases near the critical point. Given that the operating conditions are far from the critical point, this does not negatively impact.

The calculation of the specific heat capacity at constant pressure is estimated by considering the real gas equation of state, combined with corrections from the virial equation. This correction ensures that, under pressures ranging from 0.1 to 20 MPa and helium temperatures up to 1500 K[24].

The dynamic viscosity and thermal conductivity of helium-xenon (He-Xe) mixtures were calculated using Hirschfelder's method. Specifically, the viscosity calculation employed Chapman-Enskog theory, which assumes spherical molecules with no internal degrees of freedom. The thermal conductivity calculation incorporated Singh's third-order correction factor to Hirschfelder's method, accounting for higher-order effects of mass and temperature[25].

### 2.1.2 Reactor model

The reactor model includes a neutron kinetics model and a multi-channel thermal-hydraulic model.

### 2.1.2.1 Neutron kinetics

Neutron kinetics use the point reactor model, with the governing equation as

$$\begin{aligned} \frac{dP_{\text{fiss}}(t)}{dt} &= \frac{\rho(t) - \sum_{i=1}^{6}\beta_i}{\Lambda} P_{\text{fiss}}(t) + \sum_{i=1}^{6} \lambda_i C_i(t) \\ \frac{dC_i(t)}{dt} &= \frac{\beta_i}{\Lambda} P_{\text{fiss}}(t) - \lambda_i C_i(t) \quad i=1,2,\ldots,6 \\ \rho(t) &= \rho_{\text{in}}(t) + \rho_{\text{fuel}}\left(T_{\text{fuel,avg}}\right) + \rho_{\text{block}}\left(T_{\text{block,avg}}\right) \end{aligned} \quad (1)$$

where the input variables are input reactivity $\rho_{\text{in}}(t)$, average fuel temperature $T_{\text{fuel,avg}}$, and average moderator temperature $T_{\text{block,avg}}$. The output variable is the fission power $P_{\text{fiss}}(t)$. Known parameters include the fractions of six groups of delayed neutrons $\beta_{i=1\cdots 6}$, the mean neutron generation time $\Lambda$, and the decay constants $\lambda_{i=1\cdots 6}$. The intermediate variable is the concentration of delayed neutron precursors $C_{i=1\cdots 6}(t)$. The initial conditions for the differential equations are

$$\frac{dP_{\text{fiss}}(t=0)}{dt} = 0$$

$$\frac{dC_i(t=0)}{dt} = 0 \quad i = 1, 2, \ldots, 6 \quad (2)$$

$$P_{\text{fiss}}(t=0) = P_{\text{fiss},0}$$

where $P_{\text{fiss0}}$ is the initial fission power. The temperature feedback coefficients of the fuel and moderator, as well as the neutron kinetics parameters, are based on the pin-block reactor configuration from the Prometheus project [26] and are calculated using the Monte Carlo code RMC[27].

Geometrically, the horizontal and vertical cross-sections of the reactor core are shown in Figure 4. A large safety rod is positioned at the center of the core.

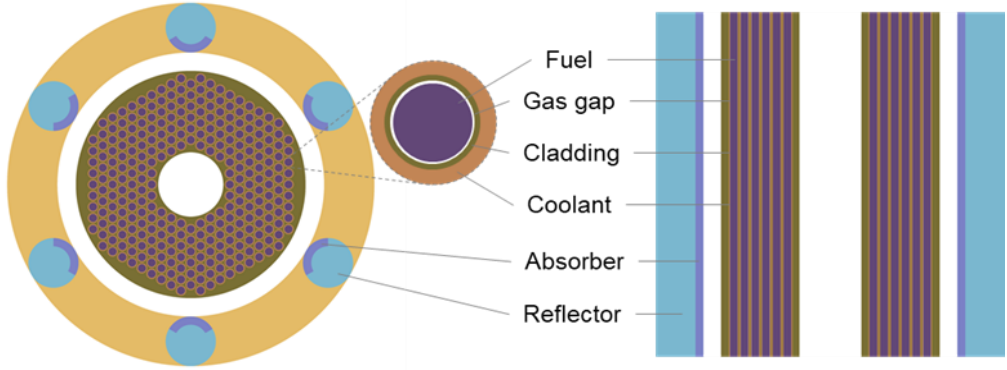

Figure 4 Horizontal and vertical cross-section of the reactor core

Regarding material settings, the core uses 90% enriched UO2 fuel, beryllium as the radial movable reflector, and Mo-47.5%Re alloy as the structural support and spectrum shift absorber.

For Monte Carlo calculations, 100,000 neutron simulations are set, with 300 iterations per generation, ignoring the results of the first 30 generations.
In the case settings, the impact of Doppler effect and thermal expansion due to temperature changes on reactivity is considered. Specifically, this includes changes in material density and geometric dimensions due to temperature, changes in absorption and fission cross-sections, and the effect of temperature on scattering cross-sections. Therefore, at each temperature point, not only are the temperature data for each nuclide considered, but geometry and material density are also reset according to the uniform expansion hypothesis to reflect macroscopic thermal expansion and contraction effects.

The temperature reactivity relationship of the fuel is fitted with a third-order polynomial, as shown in Figure 5 (a). The maximum absolute reactivity deviation of the fit is $9.54 \times 10^{-4}$. The temperature reactivity relationship of the moderator is fitted with a third-order polynomial, as shown in Figure 5 (b). The maximum absolute reactivity deviation of the fit is $1.49 \times 10^{-4}$.

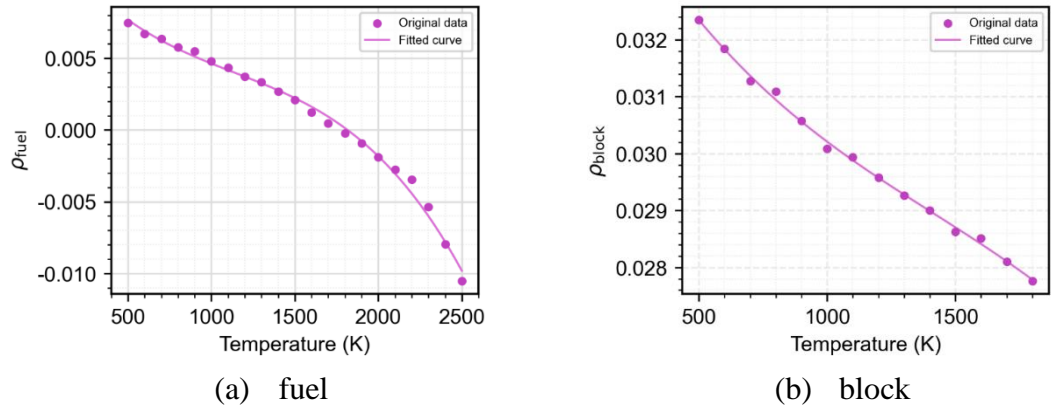

(a) fuel  (b) block

Figure 5 Temperature reactivity of the fuel and the block

2.1.2.2 Core thermal-hydraulics

The basic structural units of the core from inside to outside are fuel, gap and cladding, coolant channels, and block. At the same radial position, these units are approximately uniformly arranged in a hexagonal pattern. Thus, each ring of hexagonal units can be considered as a single channel with a uniformly distributed radial arrangement in the polar coordinate system. Multiple such single channels together form the core, with energy transfer between channels occurring through the thermal conductivity of the block. Therefore, the core can be viewed as a multi-channel structure.

This multi-channel core structure is based on the following assumptions: (1) The materials in each part of a single channel are homogeneous; (2) Heat conduction occurs only within the fuel, cladding, gap, and moderator, with no thermal contact resistance at the interfaces; (3) Only convective heat transfer occurs between the block and coolant; (4) No heat conduction occurs within the coolant channel of each single channel, and the physical parameters of the coolant are lumped radially and vary only axially; (5) The cladding and gap have negligible thermal capacity due to their much smaller thickness compared to the fuel and moderator, thus they are assumed to have only thermal resistance. The multi-channel reactor core assumption is shown in Figure 6.

The fuel interior satisfies the following partial differential equation

$$\frac{\partial(\rho_\mathrm{f} c_{\mathrm{p,f}} T_\mathrm{f})}{\partial t} = \frac{1}{r}\frac{\partial}{\partial r}\left(r k_\mathrm{f} \frac{\partial T_\mathrm{f}}{\partial r}\right) + k_\mathrm{f} \frac{\partial^2 T}{\partial z^2} + Q_\mathrm{f} \tag{3}$$

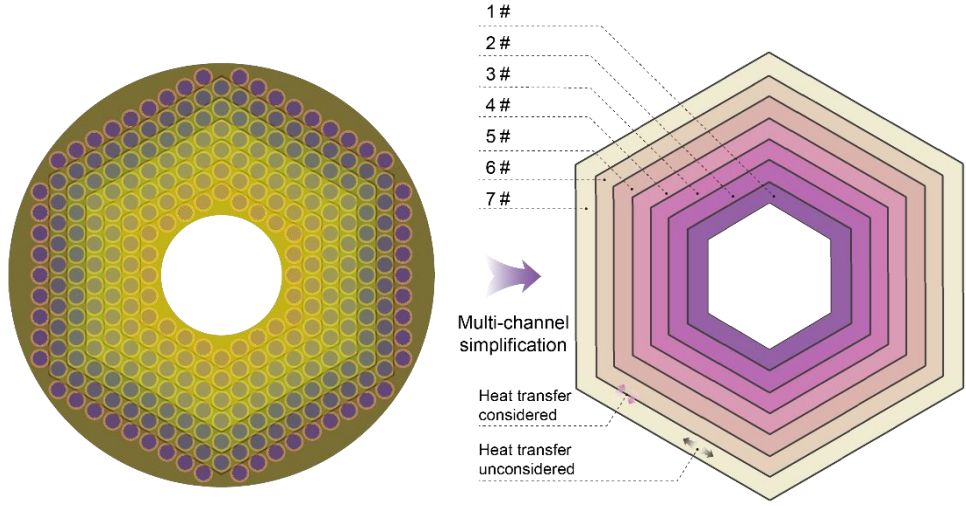

Figure 6 Multi-channel assumption for reactor core

At the boundary in the $z$ direction, the adiabatic condition is satisfied; at the boundary in the $r$ direction, it is connected to the adjacent gap region through the same temperature. Here, $\rho_\mathrm{f}$ represents the fuel density in kg/m³; $c_{\mathrm{p,f}}$ represents the specific heat capacity of the fuel in J/(kg·K); $T_\mathrm{f}$ represents the fuel temperature in K; $t$ represents time in s; $r$ represents the radial coordinate in m; $k_\mathrm{f}$ represents the thermal conductivity of the fuel in W/(m·K); $z$ represents the axial coordinate in m; $Q_\mathrm{f}$ represents the heat source term within the fuel, calculated from the output $P_{\mathrm{fiss}}$ of the neutron kinetics model and the total volume of the core fuel in W/m³.

The mass conservation equation states that the sum of the change in gas density over time and the change in mass flow rate over space at a given cross-section in the pipeline is zero, expressed as

$$\frac{\partial \rho_\mathrm{g}}{\partial t} + \frac{\partial}{\partial z}\left(\frac{\dot{m}}{A_\mathrm{p}}\right) = 0 \tag{4}$$

where, $\rho_\mathrm{g}$ represents the gas density in kg/m³; $\dot{m}$ represents the gas mass flow rate in kg/s; $A_\mathrm{p}$ represents the flow channel cross-sectional area in m².

The momentum conservation equation is expressed as

$$\frac{\partial}{\partial t}\left(\frac{\dot{m}}{A_\mathrm{p}}\right) + \frac{\partial}{\partial z}\left(\frac{\dot{m}^2}{\rho_\mathrm{g} A_\mathrm{p}^2}\right) = -\frac{\partial p_\mathrm{p}}{\partial z} - \frac{f_\mathrm{p} \dot{m}|\dot{m}|}{2 D_\mathrm{p} \rho_\mathrm{g} A_\mathrm{p}^2} - \rho_\mathrm{g} g \cos(\theta) \tag{5}$$

where on the left side, the first term represents the change in momentum over time, and the second term represents the change in momentum over space. On the right side, the first term is the momentum change due to the pressure gradient, the second term is the momentum loss due to wall friction, and the third term represents the gravity term (which is small in deep space). Here, $p_p$ represents the gas pressure in Pa; $f_p$ represents the Darcy friction factor (dimensionless); $D_p$ represents the hydraulic diameter of the pipe in meters; $\theta$ represents the angle between the gas flow direction and the positive z-axis in radians. The expression for the Darcy friction factor can be found in reference [28], and is expressed as

$$f = \begin{cases} \dfrac{64}{Re}, Re < Re_L \\ f_L + \dfrac{f_T - f_L}{Re_T - Re_L}(Re - Re_L), Re_L < Re < Re_T \\ \left(-1.8 \lg\left(\left(\dfrac{\varepsilon}{3.7 D_h}\right)^{1.11} + \dfrac{6.9}{Re}\right)\right), Re > Re_T \end{cases} \quad (6)$$

The energy conservation equation is expressed as

$$\frac{\partial (\rho_g c_p T)}{\partial t} + \frac{\partial}{\partial z}\left(\frac{\dot{m} c_p T}{A_p}\right) = Q_{in} + Q_{out} \quad (7)$$

where on the left side, the first term represents the change in the internal energy of the gas over time, and the second term represents the change in energy over space. On the right side, the two terms represent the heat transferred to the gas from the inner wall and the outer wall, respectively. Here, $c_p$ represents the specific heat capacity of the gas in J/(kg·K); $T$ represents the gas temperature in K; $Q_{in}$ represents the heat transfer from the inner wall in W; $Q_{out}$ represents the heat transfer from the outer wall in W.

The geometric characteristics of the flow channel are expressed as

$$A_p = \pi(r_{out}^2 - r_{in}^2), L = 2\pi(r_{in} + r_{out}), D_p = \frac{4 A_p}{L} \quad (8)$$

where $r_{in}$ represents the inner radius of the pipe in meters; $r_{out}$ represents the outer radius of the pipe in meters; $L$ represents the perimeter of the pipe in meters.

The heat transfer equation between the coolant and the channel wall is expressed as

$$Q_{in} = \frac{h_{in} \cdot 2\pi r_{in}(T_{in} - T_g)}{A_p}$$

$$Q_{out} = \frac{h_{out} \cdot 2\pi r_{out}(T_{out} - T_g)}{A_p} \qquad (9)$$

where $h_{in}$ represents the inner heat transfer coefficient in W/(m²·K); $h_{out}$ represents the outer heat transfer coefficient in W/(m²·K); $T_{in}$ represents the inner wall temperature in K; $T_{out}$ represents the outer wall temperature in K; $T_g$ represents the gas temperature in K.

To obtain the heat transfer coefficients, the Nusselt number is calculated using the relations from references [29,30] as follows

$$Nu_{in} = \frac{(\xi/8)RePr}{1.07 + 12.7\sqrt{(\xi/8)}\left(Pr^{2/3} - 1\right)} \theta_{in}^{-0.505 \lg \theta_{in} - 0.165}$$

$$Nu_{out} = \frac{(\xi/8)RePr}{1.07 + 12.7\sqrt{(\xi/8)}\left(Pr^{2/3} - 1\right)} \theta_{out}^{-0.505 \lg \theta_{out} - 0.165} \qquad (10)$$

$$Re = \frac{\dot{m}_p D_p}{A_p \mu}, \xi = (1.82 \log Re - 1.64)^{-2}$$

$$\theta_{in} = T_{in}/T_g, \theta_{out} = T_{out}/T_g, h_{in} = \frac{Nu_{in}\lambda}{D_p}, h_{out} = \frac{Nu_{out}\lambda}{D_p}$$

where $Nu_{in}$ represents the inner Nusselt number; $Nu_{out}$ represents the outer Nusselt number; $\theta_{in}$ represents the inner temperature ratio; $\theta_{out}$ represents the outer temperature ratio; $\lambda$ represents the thermal conductivity in W/(m·K); $\xi$ represents the friction factor; Re represents the Reynolds number; and $\mu$ represents the dynamic viscosity of the gas in Pa·s.

Finally, these physical properties are calculated using the non-ideal gas property model for helium-xenon mixtures, expressed as

$$\mu = \mu(T,p), \lambda = \lambda(T,p), c_p = c_p(T,p), Pr = Pr(T,p), \rho = \rho(T,p) \qquad (11)$$

### 2.1.3 Turbo-machinery model

The turbo-machinery model calculates the pressure and temperature changes through the compressor and turbine based on characteristic curves. The characteristic curves express the relationship between the pressure ratio, temperature ratio, dimensionless flow rate, and dimensionless rotational speed.

The pressure ratio, $p_{o2}/p_{o1}$ or $p_{o4}/p_{o5}$, represents the ratio of higher total pressure to lower total pressure. The temperature ratio, $T_{o2}/T_{o1}$ or $T_{o4}/T_{o5}$, represents the ratio of higher total temperature to lower total temperature. Subscripts 1, 2, 4, and 5 correspond to the compressor inlet, compressor outlet, turbine inlet, and turbine outlet, respectively. The subscript "o" in the characteristic curves indicates total pressure or total temperature.

The dimensionless flow rate $\dot{m}'$ and dimensionless rotational speed $N'$ are the actual flow rate and rotational speed corrected to dimensionless quantities, used to standardize performance comparisons under different operating conditions. The calculation formulas are as follows

$$\dot{m}' = \frac{\dot{m} \cdot \sqrt{\frac{T_{in} R_o}{\gamma}}}{(2r_{tip})^2 P_{in}} \tag{12}$$

$$N' = \frac{N \cdot 2r_{tip}}{\sqrt{\gamma R_o T_{in}}}$$

where $\dot{m}$ is the actual mass flow rate in kg/s; $T_{in}$ is the inlet total temperature in K; $P_{in}$ is the inlet total pressure in Pa; $R_o$ is the gas constant in J/(kg·K); $\gamma$ is the specific heat ratio; $r_{tip}$ is the impeller outer diameter; $N$ is the actual rotational speed in rpm.

Then the relationships that these four physical quantities satisfy are as follows

$$\frac{p_{o2}}{p_{o1}} = f_{prC}(\dot{m}', N'), \frac{p_{o4}}{p_{o5}} = f_{prT}(\dot{m}', N')$$
$$\frac{T_{o2}}{T_{o1}} = f_{TrC}(\dot{m}', N'), \frac{T_{o4}}{T_{o5}} = f_{TrT}(\dot{m}', N') \tag{13}$$

where $f_{prC}$ represents the compressor pressure ratio function; $f_{prT}$ represents the turbine pressure ratio function; $f_{TrC}$ represents the compressor temperature ratio function; $f_{TrT}$ represents the turbine temperature ratio function.

In dynamic analysis, the power balance among the coaxial compressor, turbine, and generator in the Brayton system is expressed as

$$P_{shaft} = P_T - P_C - P_{alt}$$
$$P_T = W \cdot (c_{p,T,in} T_{T,in} - c_{p,T,out} T_{T,out})$$
$$P_C = W \cdot (c_{p,C,out} T_{C,out} - c_{p,C,in} T_{C,in}) \tag{14}$$
$$P_{shaft} = I \cdot \omega_{shaft} \frac{d\omega_{shaft}}{dt}$$

where $P_{shaft}$ is the net shaft power; $P_T$ is the turbine power output; $P_C$ is the compressor power consumption; $P_{alt}$ is the alternator power output; $W$ is the mass flow rate; $c_{p,T,in}$ and $c_{p,T,out}$ are the specific heat capacities at the turbine inlet and outlet, respectively; $T_{T,in}$ and $T_{T,out}$ are the temperatures at the turbine inlet and outlet, respectively; $c_{p,C,in}$ and $c_{p,T,out}$ are the specific heat capacities at the compressor inlet and outlet, respectively; $T_{C,in}$ and $T_{C,out}$ are the temperatures at the compressor inlet and outlet, respectively; $I$ is the moment of inertia of the shaft; and $\omega_{shaft}$ is the angular velocity of the shaft.

### 2.1.4 Recuperator

The recuperator uses two counterflow helium-xenon gas channels connected through conductive components. In terms of control equations, the helium-xenon gas flow and heat transfer equations are the same as those within the core, and the conductive components follow a heat conduction control equation similar to that of the fuel, but without a heat source term. For boundary conditions, the outer surfaces of the two channels have adiabatic boundary conditions, and the flow boundaries are connected to adjacent components through pressure, temperature, and mass flow rate.

### 2.1.5 Cooler

The cooler is a helium-xenon flow channel similar to those in the core, with a simplified radiation boundary condition on the pipe wall, adjusted for heat dissipation area. The control equation is

$$h_{out}\left(T_{out} - T_g\right) = -\frac{S_{ts}}{S_{rs}} \varepsilon \sigma \left(T_{out}^4 - T_{space}^4\right) \tag{15}$$

where $\varepsilon$ is the emissivity of the cooler; $T_{out}$ is the wall temperature on the outside of the pipe; $T_{space}$ is the space temperature; $S_{ts}$ is the surface area of the cooler's pipe wall; $S_{rs}$ is the effective heat dissipation area of the cooler.

## 2.2 Optimization framework

This paper proposes an optimization framework to enable the reactor to transition more rapidly from cold start to rated conditions. This framework consisting of four modules: IO module, simulation module, post-processing module, and optimization algorithm module. The optimization module controls the overall optimization process. The activity diagram for this framework is shown in Figure 11.

IO module writes the control sequence files to the project directory and saves the data to disk. Simulation module uses files to perform simulations with any simulation software, obtaining changes in all variables for the current generation. Post-processing Module receives variables and calculates the performance metrics. Optimization algorithm module employs algorithms to define a multi-objective optimization problem based on these metrics and generates the control sequences for the current generation. This study employs NSGA-II[12] as the optimization algorithm.

In each iteration, the optimization module generates new control sequences, which the IO module saves for the next simulation. This iterative process continues until the optimal control sequence is achieved, ultimately yielding the optimization results.

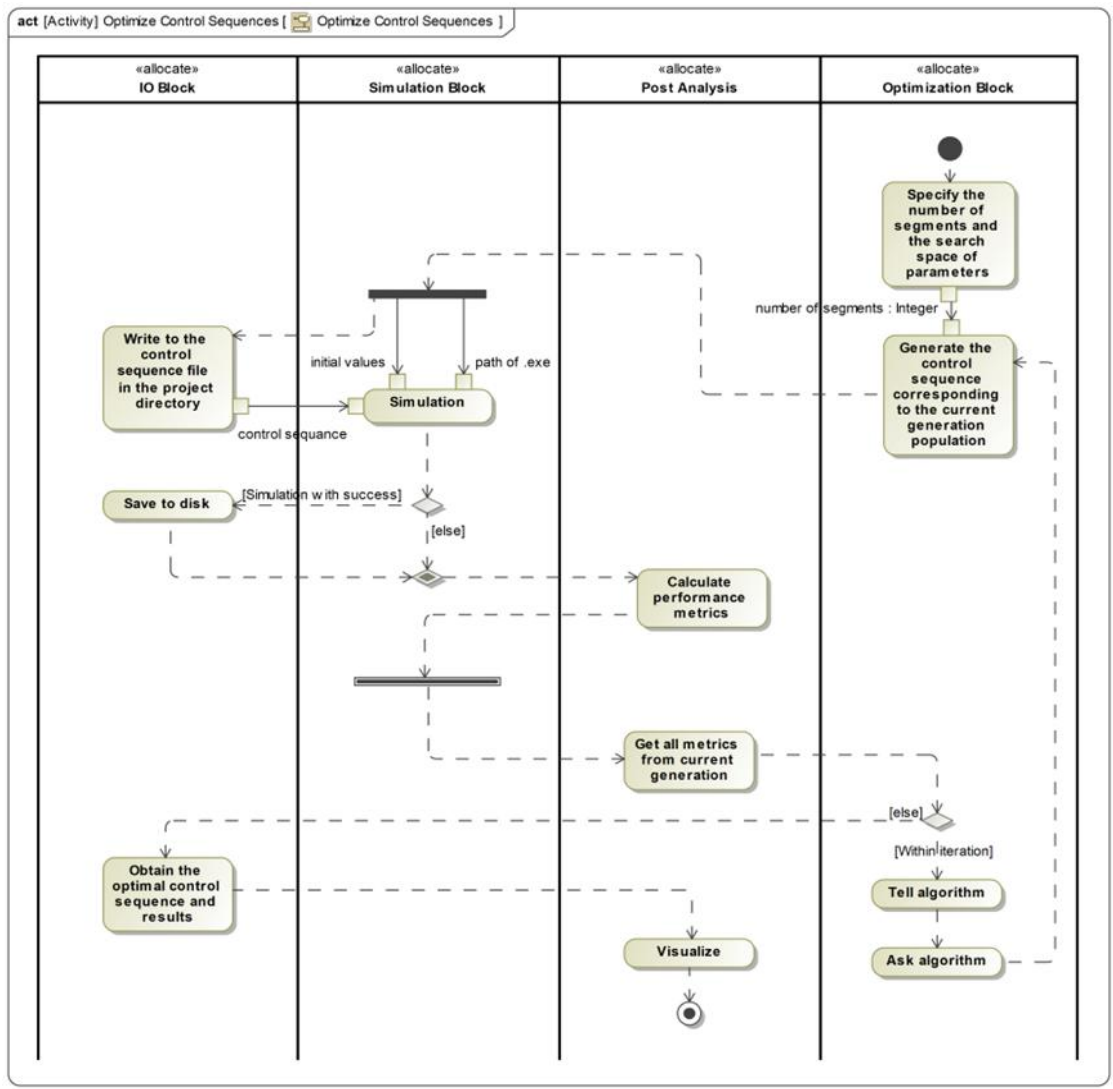

Figure 7 The framework for parameterization and optimization control sequence

# 3 Verification of models

## 3.1 Components verification

### 3.1.1 Non-ideal gas model verification

The density calculation result is shown in Figure 13 and Figure 14. As the molar mass of helium-xenon mixtures increases, the compression factor rises slowly, peaks around 30.34-61.07 g/mol, and then declines. Temperature and pressure significantly affect the compression factor; lower temperatures and higher pressures increase deviations from ideal gas behavior. At low temperatures and high molar mass, the density difference between ideal and non-ideal gases is more pronounced, requiring the mixture to be treated as a non-ideal gas for accurate density measurements.

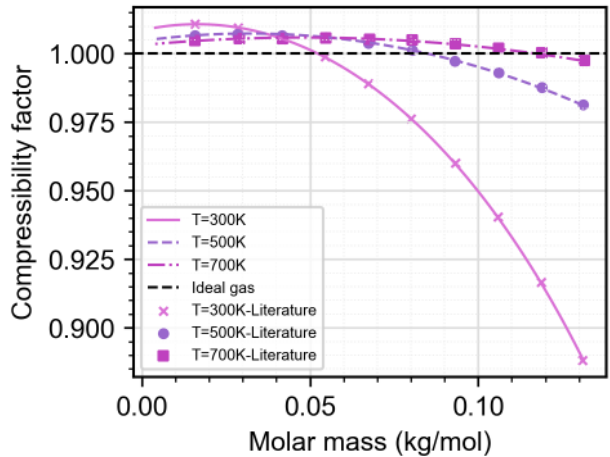

Figure 8 Effect of molar mass on the compression factor at constant pressure (2.0 MPa)

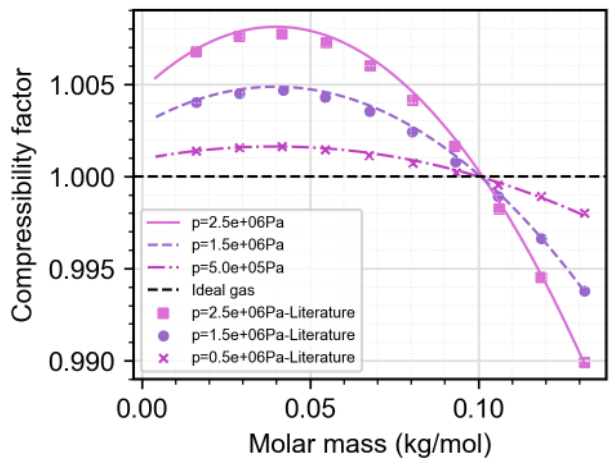

Figure 9 Effect of molar mass on the compression factor at constant temperature (600K)

The variation of molar heat capacity at constant pressure with molar mass under fixed pressure and temperature conditions is shown in Figure 10 and Figure 11, respectively. Adding xenon increases gas heat capacity, particularly at low temperatures and high pressures. As xenon proportion increases, heat capacity rises, unlike the ideal gas model's prediction. This shows gas composition significantly affects heat capacity under non-ideal conditions. Considering intermolecular interactions improves prediction accuracy in these conditions.

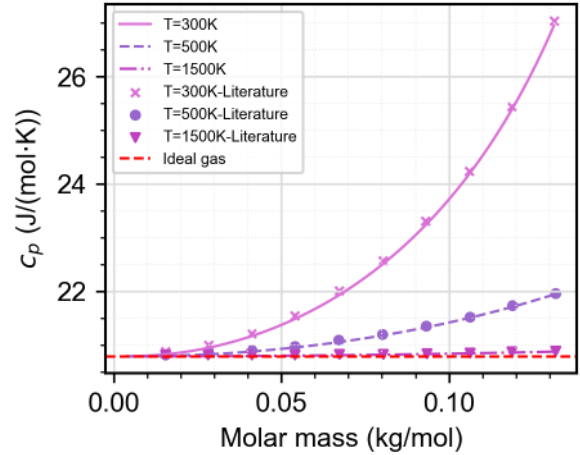

Figure 10 Molar heat capacity at constant pressure varies with molar mass at 2.0 Mpa

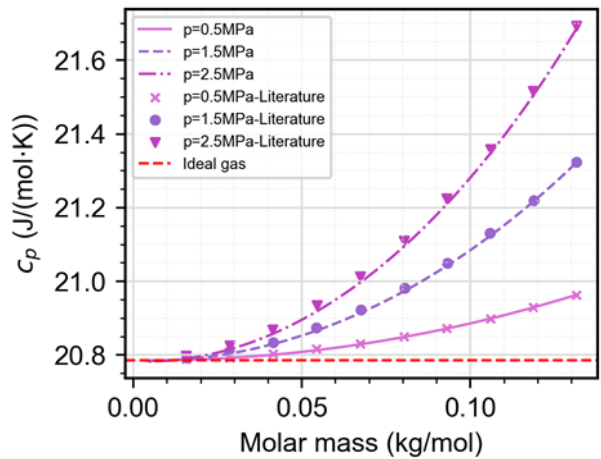

Figure 11 Molar heat capacity at constant pressure varies with molar mass at 600K

Changes in dynamic viscosity and thermal conductivity with gas molar mass are shown in Figure 12 and Figure 13, respectively. The data from the figures closely match those in reference[25]. In the temperature range of 400K to 1200K, for helium-xenon mixtures of different molecular weights, the calculated first-order dynamic viscosity values deviate from AiResearch data by no more than 0.51%. The calculated third-order thermal conductivity values deviate from AiResearch data by 2.13% to 4.40%.

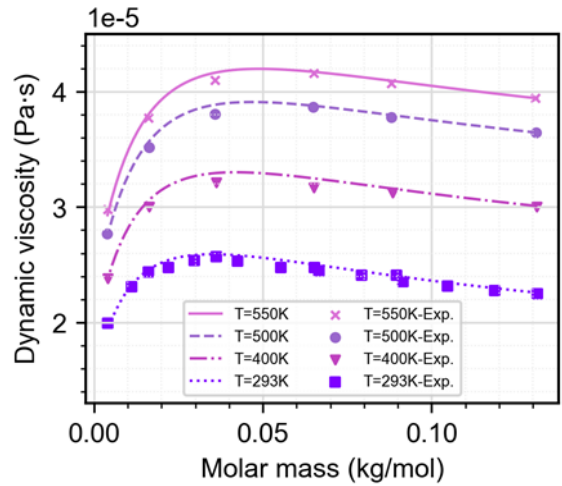

Figure 12 Dynamic viscosity with gas molar mass at different temperatures

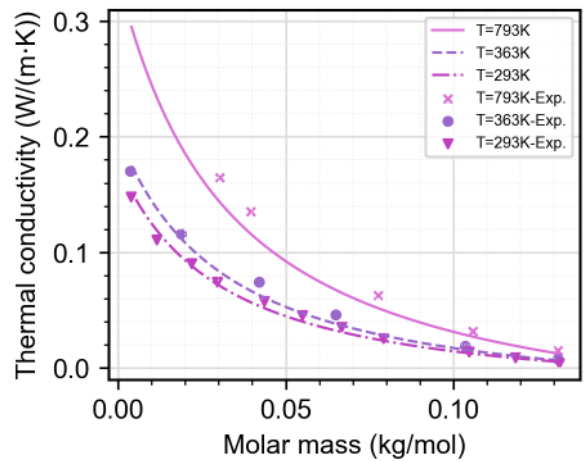

Figure 13 Thermal conductivity with gas molar mass at different temperatures

### 3.1.2 Verification of turbo-machinery model

This paper uses the Capstone C-30[31] compressor and turbine model. The characteristic curves are obtained by polynomial fitting as described in reference[31], and the compressor pressure ratio and temperature ratio curves can be seen in Figure 14 and Figure 15, respectively. The dashed line on the left represents the compressor surge boundary.

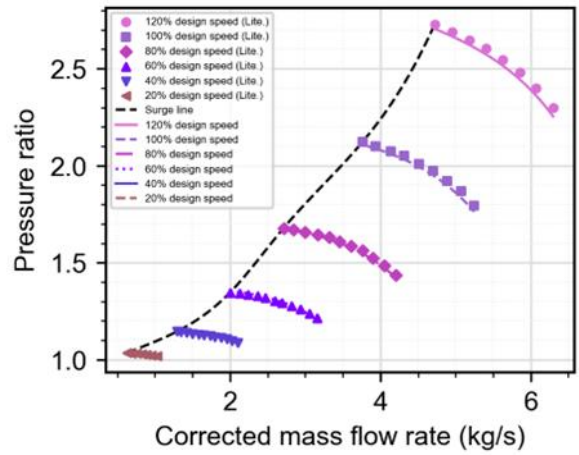

Figure 14 Compressor pressure ratio curve

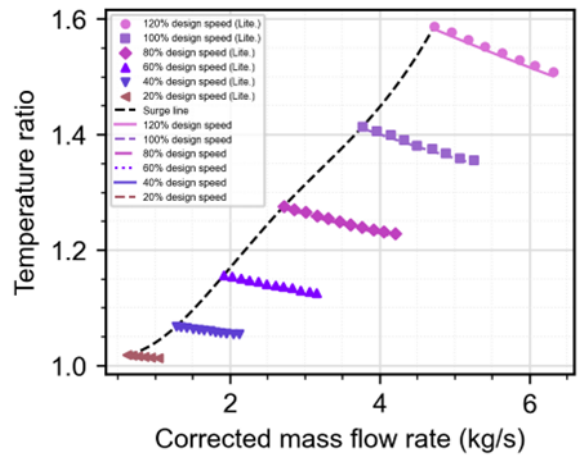

Figure 15 Compressor temperature ratio curve

The turbine pressure ratio and temperature ratio curves are shown in Figure 16 and Figure 17, respectively.

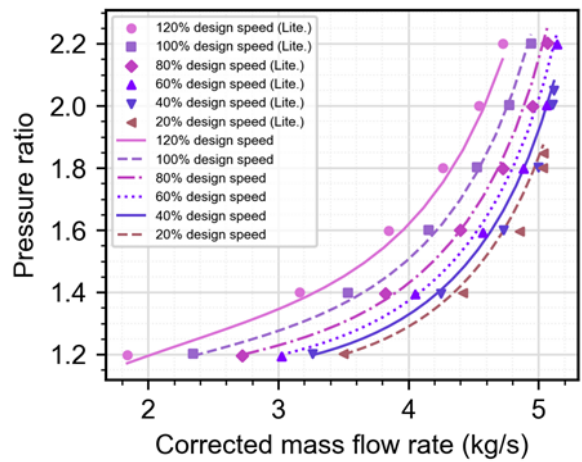

Figure 16 Turbine pressure ratio curve

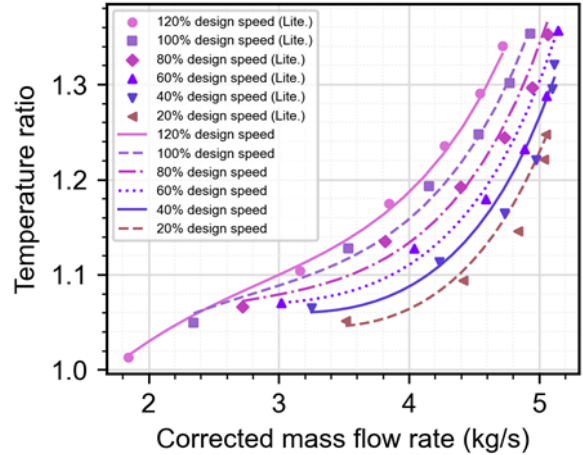

Figure 17 Turbine temperature ratio curve

## 3.2 System verification

### 3.2.1 Steady-state verification

This paper uses a reactor connected to a single Brayton system. Reference[31] from the Prometheus project's JIMO initiative describes a system with one reactor and two Brayton systems (dual-loop), but for simulation, it used one reactor and one Brayton system (single-loop). Thus, this paper uses the same configuration.

To verify the correctness of the system program, this paper compares the results obtained after the system runs transiently to steady state at rated power. The thermal line calculation results for the single-loop Brayton reactor system are from the literature, with temperature, pressure, and flow rate at various points obtained from Sandia National Laboratories' design codes FEPSIM and RxPwrSys[31]. The comparison between this paper and the literature is shown in Table 1. Parameter deviations are within 10%, with pressure-related parameters mostly around 7% and temperature-related parameters ranging from -0.8% to -9.6%. The differences may be due to:

1. Differences in core simplification assumptions. The literature uses a single average fuel rod with only axial distribution, while this paper uses a multi-channel core model with both axial and radial distributions.

2. Differences in physical property parameters. The literature assumes constant specific heat capacity for the helium-xenon mixture, while this paper updates gas properties in real-time according to the virial equation and dense gas correction theory with changing temperature and pressure.

3. Differences in component pressure drop assumptions. The literature does not consider the pressure drop in the recuperator and cooler, while this paper details the pressure drops due to friction in each component.

Given the system's complexity and that most parameter deviations are within acceptable ranges, these verification results can be considered acceptable. Although the maximum deviation is slightly higher, it remains reasonable for such a complex system.

Table 1 Comparison of steady-state thermal line results at rated conditions

|  | Reference[31,32] | Calculated | Relative error (%) |
|---|---|---|---|
| Compressor inlet temperature (K) | 319.0 | 304.8 | -4.5 |
| Compressor outlet temperature (K) | 470.0 | 427.4 | -9.1 |
| Compressor inlet pressure (MPa) | 1.50 | 1.50 | 0.0 |
| Compressor outlet pressure (MPa) | 3.00 | 2.79 | -7.0 |
| Turbine inlet temperature (K) | 1147.0 | 1111.7 | -3.1 |
| Turbine outlet temperature (K) | 908.0 | 887.8 | -2.2 |
| Turbine inlet pressure (MPa) | 2.97 | 2.74 | -7.7 |
| Turbine outlet pressure (MPa) | 1.50 | 1.51 | 0.7 |
| Cooler inlet temperature (K) | 463.0 | 418.5 | -9.6 |
| Turbine inlet pressure (MPa) | 1.50 | 1.50 | 0.0 |
| Core inlet temperature (K) | 885.0 | 877.6 | -0.8 |
| Core inlet pressure (MPa) | 3.00 | 2.78 | -7.3 |
| Mass flow rate (kg/s) | 3.10 | 3.31 | 6.8 |

### 3.2.2 Transient verification

This paper follows the startup method from reference[31], dividing the startup process into five stages. The control variables are core reactivity insertion, turbine alternator system shaft speed, and cooler outer side radiation sink temperature. The variations of these variables over time are shown in Figure 18.

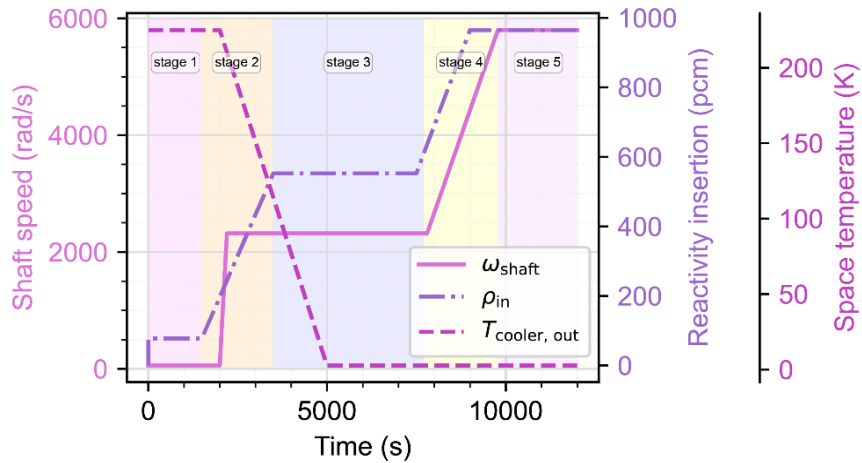

Figure 18 Control variables during the startup process

Stage 1 (0-1500 seconds): Zero power startup. Initially, the core power is only a few watts. A constant reactivity of 78 pcm is introduced and maintained for 1500 seconds, until the fission power reaches approximately 1.6 kW and the average core temperature reaches 292 K.

Stage 2 (1500-3500 seconds): Reactivity ramp insertion. During this period, reactivity increases at a constant rate, reaching 553 pcm by 3500 seconds. Around 2000 seconds, when the average core temperature reaches approximately 400 K, the turbine alternator system starts. Over the next 200 seconds, the shaft speed rises to 40% of the rated speed, initiating the Brayton system. As shown in Figure 19, the flow rate increases rapidly. At the moment of shaft speed initiation, the radiation sink temperature on the cooler's outer side begins to drop, reflecting the deployment of the cooling panels, providing a cold source for the Brayton system post-startup. Due to increased cooling power, the compressor inlet and outlet temperatures decrease to some extent.

Stage 3 (3500-7500 seconds): Low power steady state. During this period, reactivity and shaft speed remain stable, and the cooler outer side temperature continues to decrease, reaching the cosmic background radiation temperature of 2.725 K by 5000 seconds. Around 6000 seconds, the system reaches a steady state. By the end of this stage, the average fuel temperature is 618 K, fission power is 45 kWt, core inlet and outlet enthalpy rise is 43 kWt, and generator output power is 10 kWe.

Stage 4 (7500-9780 seconds): Transition to full power. Starting at 7500 seconds, reactivity increases at a constant rate by 412 pcm, ending at 9000 seconds. Shaft speed begins to ramp up at 7800 seconds, reaching 100% rated speed by 9780 seconds. Reactor power increases approximately linearly, as shown in Figure 20.

Stage 5 (after 9780 seconds): Full power steady state. Around 10,000 seconds, the Brayton system reaches full power steady state. At this point, the average fuel temperature is 1056 K, core outlet temperature is 1066 K, and core outlet pressure is 2.72 MPa. The core fission power is 416 kWt, the core inlet and outlet enthalpy rise is 390 kWt, and the generator output power is 98 kWe, with an efficiency of 25.13%.

The comparison of core power during the startup process with that in reference[31] is shown in Figure 20. It can be seen that the core power trends are similar, but there are some differences during the second stage of reactivity ramp insertion and the final full power steady state. In this work, the first power increase of the core occurs earlier compared to the reference, due to the higher initial power of the core at the beginning of the startup, leading to greater power changes during the period of constant reactivity insertion. At the final full power steady state, the higher total reactivity insertion in this work results in a higher steady state power, with the core outlet temperature closer to the Brayton system loop's design value of 1144 K.

The comparison of turbine inlet temperature with the data from literature[31] is shown in Figure 21. It can be observed that, in the second phase, the temperature increase rate is slower and the peak temperature is lower than the literature data. This is because the time taken for the shaft speed to rise from 0 to 40% was 200 seconds in this study, compared to 100 seconds in the literature. The purpose of this longer duration was to reduce the stiffness of the numerical equations and improve the solving speed. Consequently, more heat accumulated in the core before the shaft speed increased rapidly in the literature, leading to a faster and higher temperature rise at the core outlet, which is the turbine inlet, after the rapid shaft speed startup.

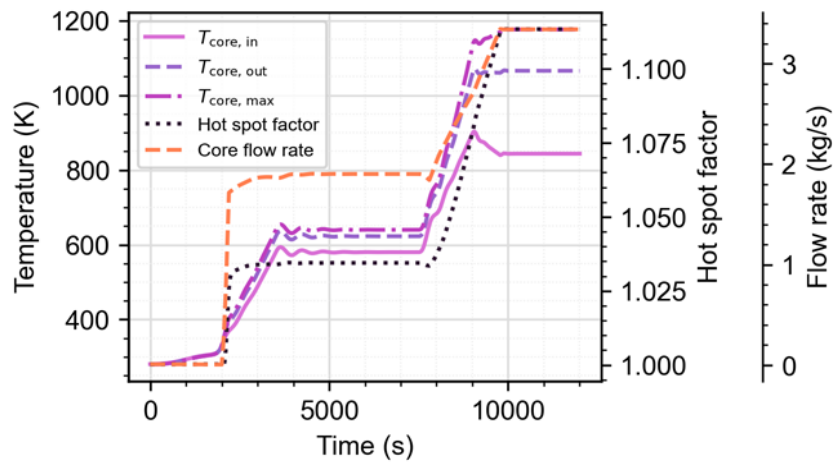

Figure 19 Changes in core temperature indicators and flow rate during the startup

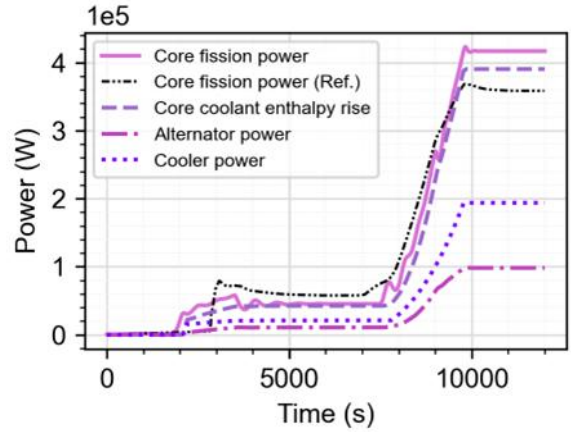

Figure 20 Power changes in various parts of the Brayton system

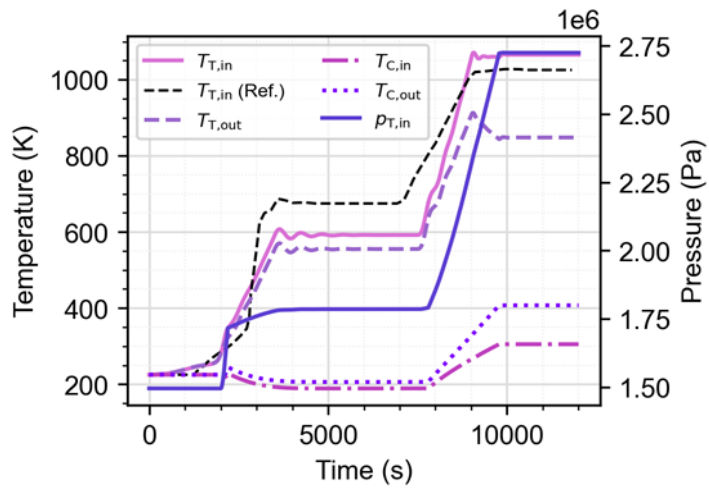

Figure 21 Temperature changes at various points in the Brayton system and turbine inlet pressure changes

# 4 Results and discussion

## 4.1 Poblem setup and solution

During the startup process, the low-power steady-state phase lasts too long, resulting in low overall startup efficiency. To address this issue, this paper optimizes the control sequences of variables such as reactivity insertion, TAC system shaft speed, and cooling system background temperature. The goal is to accelerate the system's achievement of rated power and improve startup efficiency. The optimization involves parameterizing the control sequences, setting performance evaluation metrics, and establishing the algorithmic workflow.

First, the method for parameterizing the control sequences is illustrated in Figure 22. For the shaft speed control sequence, the parameterized control points are 1# through 4#. Point 1# marks the shaft speed start time $t_1$, with the parameter being the startup time. Point 2# marks the end of the first increase in shaft speed $\omega_2$, with the parameter being the shaft speed percentage at this endpoint. The time interval between points 1# and 2# is fixed at 200 seconds to reflect the rapid startup process in a real scenario while ensuring numerical stability. Point 3# marks the start of the second shaft speed increase, with parameters for the start time $t_3$ and corresponding shaft speed $\omega_3$. Point 4# marks the point at which the shaft speed reaches its rated value, with the parameter being the corresponding time $t_4$.

For the reactivity insertion control sequence, the parameterized control points are 5# to 8#. Point 5# marks the end of the zero-power startup phase and the beginning of the first reactivity ramp insertion, with parameters for the corresponding time $t_5$ and constant reactivity $\rho_5$ during the zero-power startup. Point 6# is the end of the first reactivity ramp insertion, with parameters for time $t_6$ and reactivity $\rho_6$. Point 7# is the start of the second reactivity ramp insertion, with parameters for time $t_7$ and reactivity $\rho_7$. Point 8# marks the end of reactivity insertion during the startup process, with parameters for the corresponding time $t_8$.

For the control sequence of the cooler external temperature, the parameterized control points are 9# and 10#. Point 9# is the start of the cooler panel deployment, with the parameter being the corresponding time $t_9$. Point 10# is the end of the deployment, with the parameter being the corresponding time $t_{10}$.

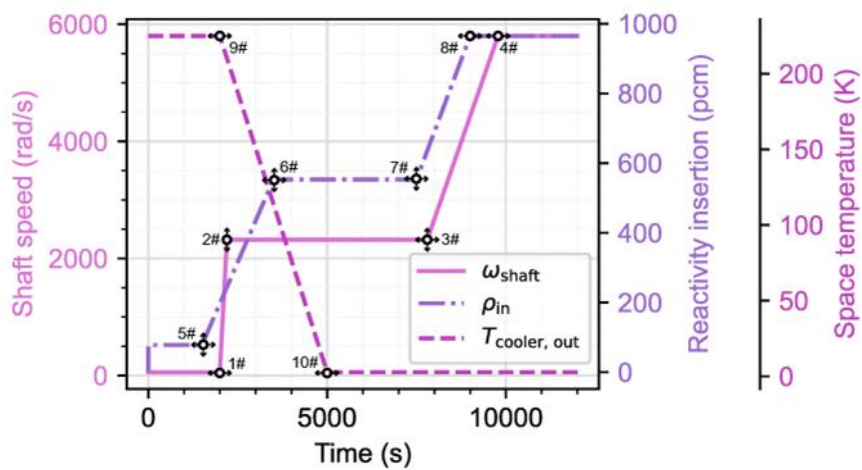

Figure 22 Parameterizing control sequences via movable points

Regarding performance metrics, the focus is on startup efficiency, considering both power and temperature aspects. On one hand, the sooner the fission power first reaches 400 kW, the better. On the other hand, the sooner the core outlet temperature first reaches 1000 K, the better. Therefore, the objective function can be expressed as follows.

$$f_1(\mathbf{x}) = t_{power\_400kW} \tag{16}$$

$$f_2(\mathbf{x}) = t_{temp\_1000K} \tag{17}$$

where vector $\mathbf{x} = (t_1, \omega_2, t_3, \omega_3, t_4, t_5, \rho_5, t_6, \rho_6, t_7, \rho_7, t_8, t_9, t_{10})$ representing control sequence parameters

The primary constraint is time, which must increase sequentially. Therefore, the optimization problem for the control sequence can be described as

$$\mathbf{X} = \arg\min_{\mathbf{x}}(f_1(\mathbf{x}), f_2(\mathbf{x}))$$

$$\text{subject to} \begin{cases} t_1 < t_3 < t_4 \\ t_5 < t_6 < t_7 < t_8 \\ t_9 < t_{10} \\ \mathbf{X}_{min} \leq \mathbf{X} \leq \mathbf{X}_{max} \end{cases} \tag{18}$$

Then, the system analysis program and the optimization framework are utilized to optimize the startup control of the He-Xe cooled space reactor system. By shortening the control time of reactivity, shaft speed and cooler background temperature, the time to reach rated power is reduced by approximately 1260 seconds, and the turbine inlet temperature reaches its rated value about 1980 seconds earlier. The detailed results are as follows.

The control strategy underwent 20 generations of evolution, with a population size of 100 per generation. Each individual in the population underwent a 10,000-second startup simulation. For successful simulations, the average CPU time was approximately 130 to 150 seconds, resulting in a total CPU time exceeding 75 hours. The optimal values of the two performance metrics per generation during the optimization process are shown in Figure 23. In the first six generations, all simulations failed due to excessive stiffness of the ordinary differential equations. From the seventh generation onward, the corresponding ODE system could be solved. Significant improvements in both performance metrics were observed after the 10th and 13th generations. The Pareto front of the multi-objective optimization has four dominant solutions, as shown in Figure 24. The control case, where power and temperature first exceed their thresholds at 8520 seconds and 6570 seconds respectively, was selected for analysis.

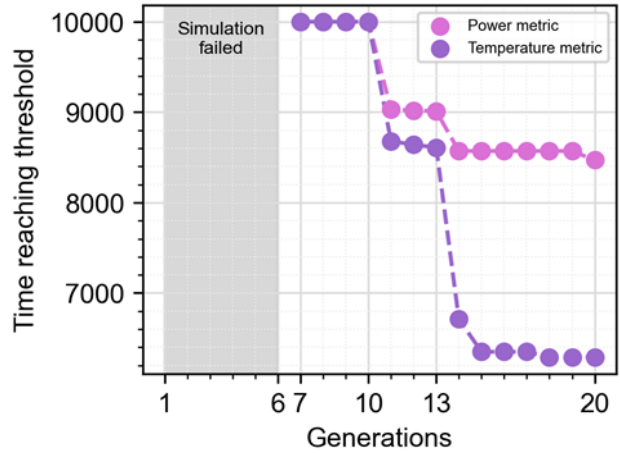

Figure 23 Changes in performance metrics with each generation

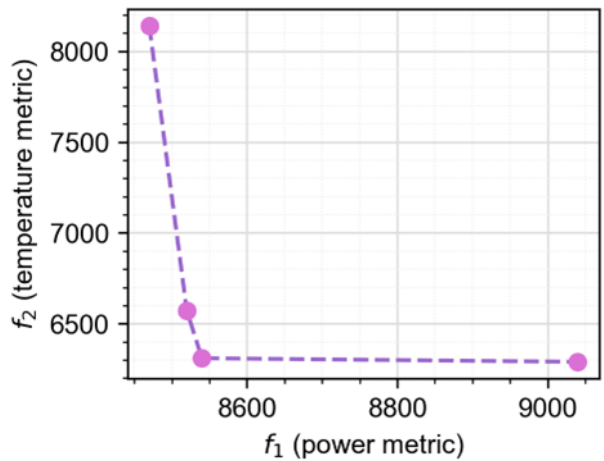

Figure 24 Pareto front of control sequence optimization

## 4.2 Optimized control variables

The time comparisons for each control stage are shown in Table 2. It is evident that the control duration for different variables significantly decreases at different stages. Notably, the duration for reactivity introduction control shows a remarkable reduction, with the time decreasing by 1169.0 seconds from stage 7# to stage 8#, a reduction of 77.9%.

Table 2 Control variable comparison

| Control variable | Phase | Original time (s) | Optimized time (s) | Shortened time (s) | Shortened percentage (%) |
|---|---|---|---|---|---|
| Shaft speed | Start to 1# | 2000.0 | 1541.6 | 458.4 | 22.9 |
|  | 1# to 2# | 200.0 | 200.0 | 0.0 | 0.0 |
|  | 2# to 3# | 5600.0 | 4631.0 | 969.0 | 17.3 |

|  | 3# to 4# | 1980.0 | 2050.3 | -70.3 | -3.6 |
| --- | --- | --- | --- | --- | --- |
| Reactivity insertion | Start to 5# | 1500.0 | 1128.5 | 371.5 | 24.8 |
|  | 5# to 6# | 2000.0 | 1749.8 | 250.2 | 12.5 |
|  | 6# to 7# | 4000.0 | 3302.7 | 697.3 | 17.4 |
|  | 7# to 8# | 1500.0 | 330.96 | 1169.0 | 77.9 |
| Cooler temperature | Start to 9# | 2000.0 | 1638.8 | 361.2 | 18.1 |
|  | 9# to 10 # | 3000.0 | 3294.5 | -294.5 | -9.8 |

The control curves for reactivity, shaft speed, and cooler background temperature are shown in Figure 25 to Figure 27, respectively. For reactivity control, the initial cold startup exhibits higher reactivity, transitioning to a low-slope ramp reactivity insertion during the low-power steady state, with a steeper slope at the final ramp insertion. Shaft speed control points occur earlier than before optimization, with the low-power steady state transitioning to a low-slope shaft speed increase. The cooler background temperature control shows minimal difference before and after optimization, indicating it is not a major factor affecting startup process efficiency.

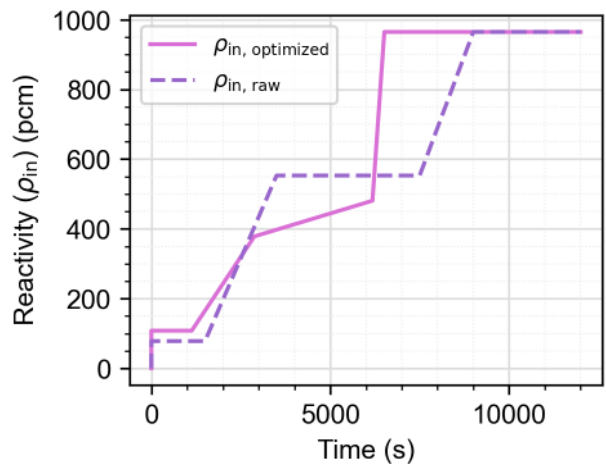

Figure 25 Comparison of reactivity insertion control before and after optimization

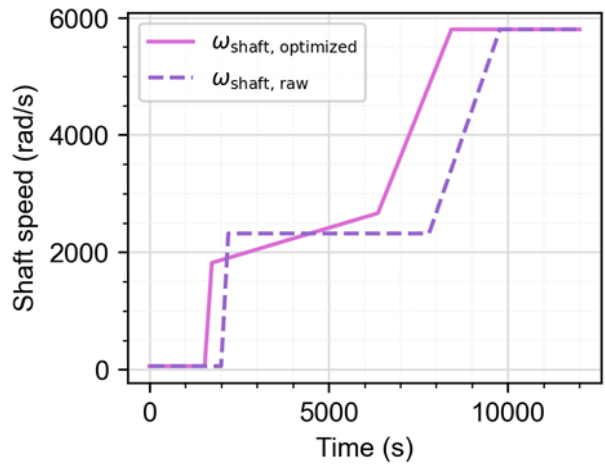

Figure 26 Comparison of shaft speed control before and after optimization

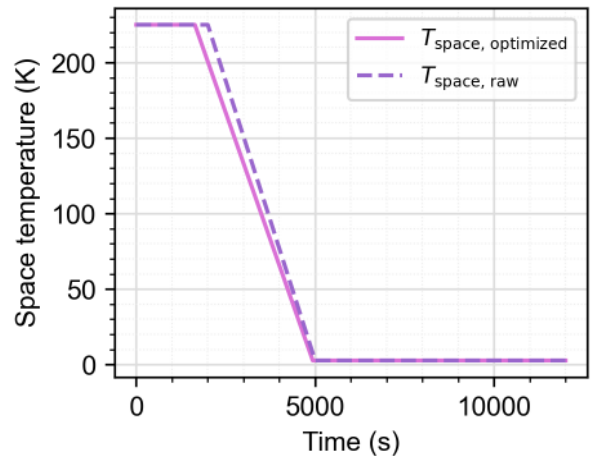

Figure 27 Comparison of cooler temperature control before and after optimization

## 4.3 Discussions

The effects of the optimized control sequences are reflected in the speed at which power and temperature rise from a cold state to rated conditions.

### 4.2.1 Fission power

The variation of fission power over time is shown in Figure 28. It can be observed that the reactivity insertion around 2000 seconds caused larger power fluctuations. Around 6000 seconds, rapid reactivity insertion resulted in a brief, sharp increase in power, but due to negative reactivity feedback, there was a significant power decrease in the following 500 seconds. Subsequently, as the power approached the rated level, the slope of the fission power increase tended to follow the changes in shaft speed.

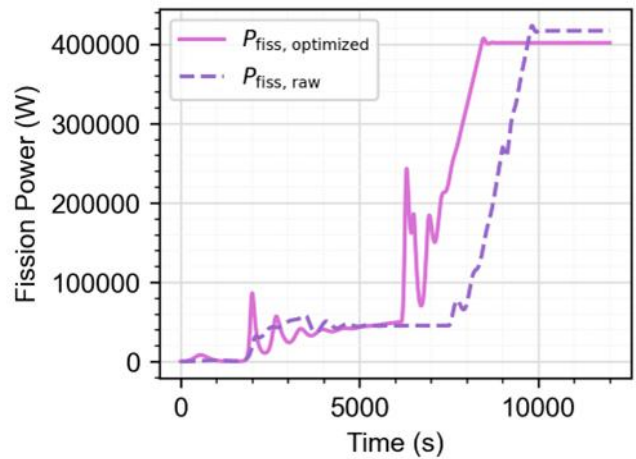

Figure 28 Comparison of cooler temperature control before and after optimization

## 4.2.2 Turbine inlet temperature

The turbine inlet temperature is shown in Figure 29. Due to the reduced reactivity at the end of the first reactivity ramp insertion compared to the pre-optimization stage, the temperature is lower around 4000 seconds. Subsequently, around 6000 seconds, the temperature rises rapidly following the swift reactivity insertion, overshoots, and fluctuates before gradually stabilizing to the rated inlet temperature. Overall, the turbine inlet temperature tends to follow the control trend of reactivity.

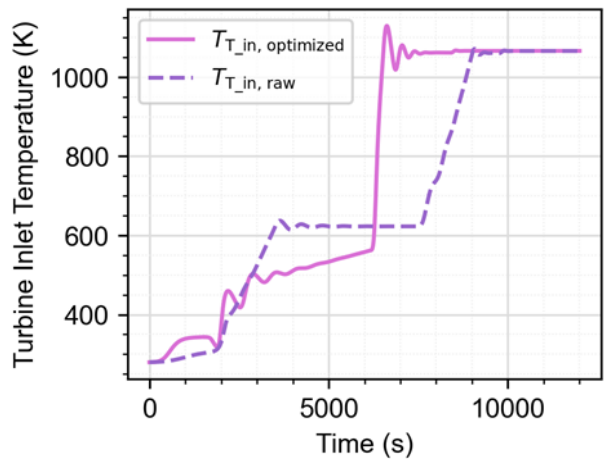

Figure 29 Comparison of turbine inlet temperature control

Additionally, the maximum core temperature, which is critical to the integrity of the core, is illustrated in Figure 30. Overall, the trend is similar to that of the turbine inlet temperature, and the peak temperature remains nearly the same as before optimization. However, the rapid temperature changes in the core caused by reactivity insertion may result in significant temperature gradients in different regions. This can lead to uneven thermal expansion, potentially causing material cracking, deformation, or structural failure.

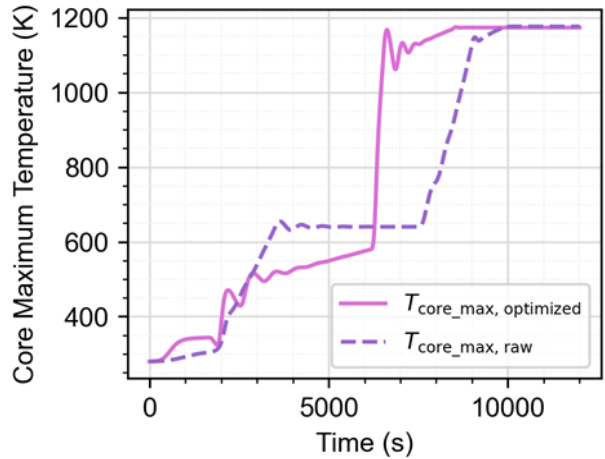

Figure 30 Comparison of core max temperature before and after optimization

### 4.2.3 Net power provided by storage batteries

The comparison of the generator's net power output over time before and after optimization, as shown in Figure 31, demonstrates the following advantages of the optimized control strategy:

Firstly, although the time to reach energy balance is longer after optimization ( $t_{optim} = 76.6\,\text{s}$ , $t_{original} = 54.6\,\text{s}$ ), the peak external power demand is significantly reduced. The minimum net power before optimization is -201.3 W, while after optimization it is reduced to -145.3 W. This indicates that the optimized strategy effectively reduces the instantaneous power demand from the battery, alleviating the burden on the energy storage system.

Secondly, the total external energy demand is decreased. The external energy demand before optimization is $E_{original} = -9264.2\text{J}$ , while after optimization it is $E_{optim} = -7673.3\text{J}$ . This demonstrates that the optimized control strategy improves energy efficiency by reducing the total energy consumption during the startup process.

Lastly, the optimized strategy enables energy self-sufficiency at a lower power level, which is crucial for space missions with limited battery capacity, improving the system's energy management efficiency.。

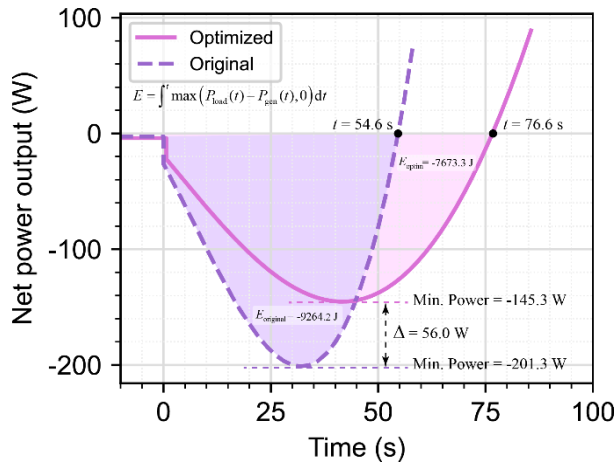

Figure 31 Comparison of the net power output of the generator over time before and after optimization

## 5 Conclusions

This paper presents the development and optimization of a Space Nuclear Power System (SNPS) with a Closed Brayton Cycle (CBC), focusing on dynamic modeling and control optimization. A comprehensive simulation code NuHeXSys was developed, integrating non-ideal gas properties, multi-channel reactor thermal-hydraulics, and turbo-machinery models. Through the use of an evolutionary algorithm, the system's startup control sequence was optimized, significantly improving startup performance. Key findings and quantitative conclusions include:

1. A detailed model for the SNPS was developed, incorporating helium-xenon gas behavior and a multi-channel reactor core. The model's steady-state and transient responses were verified, with parameter deviations falling within acceptable limits (most under 10%), confirming its accuracy in predicting system behavior under various conditions.

2. Using a multi-objective optimization algorithm (NSGA-II), the startup time of the system was reduced by approximately 1260 seconds, while the turbine inlet temperature reached rated conditions around 1980 seconds earlier. These improvements in control sequences enhanced the system's efficiency and operational stability.

3. The optimized control strategy decreased the total energy demand during startup, reducing the external energy requirement from 9264.2 J to 7673.3 J, indicating a 17% improvement in energy consumption, while reducing the burden on the storage batteries by lowering peak power demand by over 27%.

The proposed optimization method is limited in its ability to account for the full mission lifecycle priorities and multiple conflicting objectives, such as long-term reliability and resource constraints. Future research could focus on integrating multi-objective optimization across different mission phases to enhance overall system performance and adaptability.

## Declaration of competing interest

The authors declare that they have no known competing financial interests or personal relationships that could have appeared to influence the work reported in this paper.

## Acknowledgements

The authors thank the National Key Laboratory of Nuclear Reactor Technology，Nuclear Power Institute of China for their support.

## CRediT

**Chengyuan Li**: Conceptualization, Methodology, Software, Formal analysis, Investigation, Writing - Original draft preparation. **Leran Guo**: Data curation, Validation, Writing - Original draft preparation. **Jian Deng**: Supervision, Project administration. **Shanfang Huang**: Supervision, Resources, Writing - Review & Editing.

## Funding sources

This research did not receive any specific grant from funding agencies in the public, commercial, or not-for-profit sectors.